\documentclass{PoS}
\usepackage{cite}

\title{$B_s \to \mu^+ \mu^-$ in the SM and $\bar{B}\to X_s \gamma$ in the 2HDM at NNLO in QCD}

\ShortTitle{$B_s \to \mu^+ \mu^-$ in the SM and $\bar{B}\to X_s \gamma$ in the 2HDM at NNLO in QCD}

\author{\speaker{Thomas Hermann}
       \thanks{Preprint number: TTP13-041, SFB/CPP-13-101}\\
        Institut f\"ur Theoretische Teilchenphysik\\
        Karlsruhe Institute of Technology (KIT)\\
        D-76128 Karlsruhe, Germany
	KIT\\
        E-mail: \email{t.hermann@kit.edu}}

\abstract{In this contribution the QCD next-to-next-to-leading order (NNLO) matching coefficients to the branching ratio $B_s \to \mu^+ \mu^-$ in the Standard Model (SM) and
to $\bar{B}\to X_s \gamma$ in the Two Higgs Doublet Model (2HDM) are discussed. In both cases a three-loop matching between the full and effective theory
was performed, which leads to a significant reduction of the scale uncertainty.}

\FullConference{11th International Symposium on Radiative Corrections (Applications of Quantum Field Theory to Phenomenology) \\
                 22-27 September 2013\\
                 Lumley Castle Hotel, Durham, UK}

\newcommand{\gammas}{\Gamma^{\raisebox{0.2mm}{$\scriptstyle s$}}}
\newcommand{\mhplus}{M_{H^+}}
                 
\begin{document}

\section{Introduction}

Together with direct searches at the LHC, rare B meson decays are very important
for the search of physics beyond the Standard Model (SM).
Therefore, it is necessary to provide precise theory predictions to those decays.

The decay $B_s \to \mu^+ \mu^-$ yields important constraints on extensions of the SM.
Recently, LHCb and CMS have provided first measurements of the branching ratio~\cite{Aaij:2012nna,Aaij:2013aka,Chatrchyan:2013bka}
and their combined result reads~\cite{CMS.LHCb.average:2013xxx}
\begin{equation}
  \overline{{\mathcal B}}(B_s \to \mu^+ \mu^-) 
  = \left(\, 2.9 \pm 0.7 \right) \times 10^{-9}\, .
\end{equation}
Previous upper limits can be found in
Refs.~\cite{Abazov:2010fs,Aaltonen:2011fi,Chatrchyan:2012rga,Aaij:2012ac,Aad:2012pn}.
In the future a significant reduction of the experimental uncertainties is expected.
On the theory side the leptonic decay constant $f_{B_s}$ was the dominant uncertainty in the last years.  
Recent progress in the determination of $f_{B_s}$ from lattice 
calculations~\cite{Dimopoulos:2011gx,McNeile:2011ng, Bazavov:2011aa,
Bernardoni:2012fd, Carrasco:2012ps, Dowdall:2013tga} provides a motivation for
improving the perturbative ingredients, in particular the two-loop
electroweak~\cite{Bobeth:2013tba} and the three-loop QCD corrections~\cite{Hermann:2013kca}.
In Section~\ref{sec::Bsmumu} some aspects of the three-loop matching are discussed, the full discussion
can be found in~\cite{Hermann:2013kca}.

The inclusive decay $\bar{B}\to X_s \gamma$ also provides very strong constraints on physics beyond the SM.
Especially in the Two Higgs Doublet Model (2HDM) of type-II, where $\bar{B}\to X_s \gamma$
gives one of the highest exclusion limits for the charged Higgs boson mass $\mhplus$.
Therefore it is worth to calculate the three-loop QCD corrections to the corresponding Wilson coefficients in the 2HDM, which was done in
Ref.~\cite{Hermann:2012fc} and is shortly described in Section~\ref{sec::bsgamma}.
Together with all the other NNLO QCD ingredients~\cite{Misiak:2006ab,Misiak:2006zs} obtained for the SM prediction of ${\mathcal B}(\bar{B}\to X_s \gamma)$,
it is now possible to give also for the 2HDM a consistent prediction at NNLO in QCD.

\section{\label{sec::Bsmumu}$B_s \to \mu^+ \mu^-$ in the SM at NNLO in QCD}

A convenient framework for calculating the branching ratio $B_s \to \mu^+ \mu^-$,
is an effective theory derived from the SM by integrating out all heavy particles like the top quark,
the Higgs boson and the massive electroweak bosons.
The relevant effective Lagrangian for $B_s \to \mu^+ \mu^-$ reads
\begin{eqnarray}
  {\mathcal L}_{\rm eff} &=& {\mathcal L}_{\rm QCD \times QED}\mbox{(leptons and five light
    quarks)} + N \sum_n C_n Q_n ~+~ {\rm h.c.}\,,  
  \label{eq::leff}
\end{eqnarray}
with $N = V_{tb}^* V_{ts} G_F^2 M_W^2/\pi^2$. The necessary operators are
\begin{eqnarray}
 Q_A &=& (\bar{b} \gamma_{\alpha} \gamma_5 s)(\bar{\mu} \gamma^{\alpha} \gamma_5 \mu) \,,  \\
 Q_S &=& (\bar{b} \gamma_5 s)(\bar{\mu} \mu)\,,\nonumber\\
 Q_P &=& (\bar{b} \gamma_5 s)(\bar{\mu} \gamma_5 \mu) \,. \nonumber
\end{eqnarray}
In the SM only the Wilson coefficient $C_A$ matters, because
contributions from $C_S$ and $C_P$ to the branching ratio are suppresed by $M_{B_s}^2/M_W^2$.
In this case, a formula for the measured average time-integrated branching ratio~\cite{DeBruyn:2012wk} reads
\begin{equation}
  \overline{\mathcal B}(B_s\to \mu^+ \mu^-) = 
  \frac{|N|^2 M_{B_s}^3 f_{B_s}^2}{8 \pi\, \gammas_H}
  \, \beta \, r^2 | C_A (\mu_b) |^2 
   ~+~ {\mathcal O}(\alpha_{em})\,,
 \label{eq::BR}
\end{equation}
with $r = \frac{2 m_{\mu}}{M_{B_s}}$ and $\beta = \sqrt{1-r^2}$. $\gammas_H$ is the total width of the
heavier mass eigenstate.

\begin{figure}[t]
  \begin{center}
    \includegraphics[width=\textwidth]{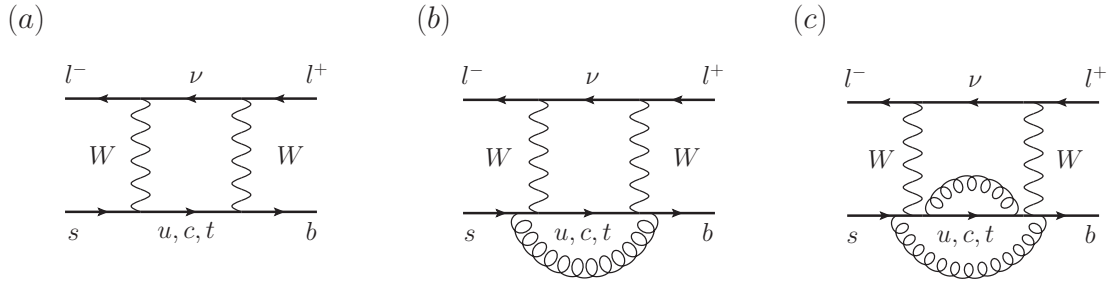}
    \caption{\label{fig::w_boxes} Sample $W$-boson box diagrams contributing to $C_A$.}
  \end{center}
\end{figure}

\begin{figure}[t]
  \begin{center}
    \includegraphics[width=\textwidth]{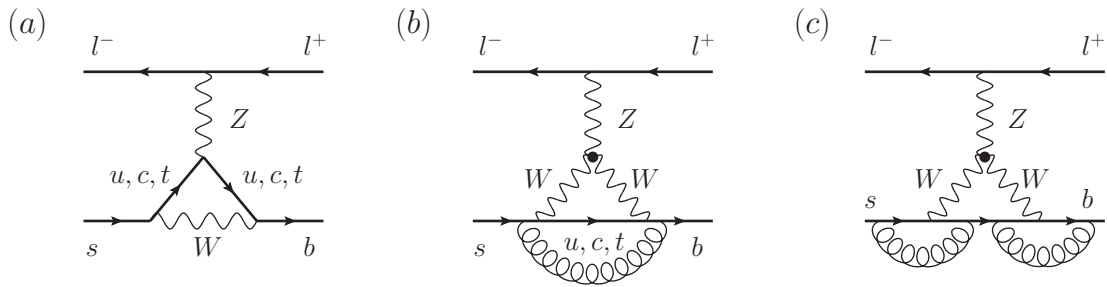}
    \caption{\label{fig::z_penguin} Sample $Z$-boson penguin diagrams contributing to $C_A$.}
  \end{center}
\end{figure}

In the SM there are two types of diagrams contributing to $C_A = C_A^{W} + C_A^{Z}$, 
the W-boson box and the Z-boson penguin diagrams (see
Fig.~\ref{fig::w_boxes} and Fig.~\ref{fig::z_penguin}),
which are discussed in the
following.
The one-loop contribution to $C_A$ has been calculated for the first time in Ref.~\cite{Inami:1980fz}
and two-loop QCD correction can be found in
Refs.~\cite{Buchalla:1992zm,Buchalla:1993bv,Misiak:1999yg,Buchalla:1998ba}.

Sample diagrams to $C_A^{W}$ at one-, two and three-loop order
are shown in Fig.~\ref{fig::w_boxes}.
Since the up and charm quark masses are set to zero, $C_A^{W}$ can be written as
\begin{eqnarray}
  C_A^{W} &=& C_A^{W,t} - C_A^{W,c}\,,
  \label{eq::CAW_split}
\end{eqnarray}
due to the unitarity of the CKM matrix.

The matching can be performed in two different ways, in $d=4-2\epsilon$ and in $d=4$ dimensions.
In the first approach we set all light quark masses to zero, which leads to spurious infrared divergences in $\epsilon$ in
the full and effective theories which cancel out in the matching procedure.
Due to the presence of those additional poles at intermediate steps, it is necessary to introduce an evanescent operator
\begin{eqnarray}
  Q^E_A &=& (\bar{b} \gamma_{\alpha_1} \gamma_{\alpha_2} \gamma_{\alpha_3}
  \gamma_5 s)(\bar{\mu} \gamma^{\alpha_3} \gamma^{\alpha_2} \gamma^{\alpha_1}
  \gamma_5 \mu) - 4 \, Q_A\,,
\end{eqnarray}
which vanishes in $d=4$ dimensions. With the mixing of $Q^E_A$ into $Q_A$, the Wilson coefficient
of the evanescent operator gives a contribution to the Wilson coefficient $C_A$.

In the second approach we introduce small masses for the strange and bottom quark as regulators for
the infrared divergences. The matching of the full and effective theory can be performed in $d=4$ dimensions,
so without contributions from evanescent operators.
After the matching it is possible to take the limits $m_s \to 0$ and $m_b \to 0$.
We have used both methods and have obtained identical results for $C_A$.

In the SM three-loop vacuum integrals with two different mass scales have to be computed.
Some classes of Feynman diagrams of this kind are known (e.g., Ref.~\cite{Grigo:2012ji}),
nevertheless we follow the same strategy as in Ref.~\cite{Misiak:2004ew}.
We expand the integrals in the limit $M_W\ll m_t$ and $M_W\approx m_t$. A combination
of those expansions gives a very good approximation to the exact three-loop result and is sufficient
for all practical purposes.
For the calculation we used the programs {\tt QGRAF}~\cite{Nogueira:1993ex} to generate the Feynman diagrams, {\tt q2e}
and {\tt exp}~\cite{q2eexp} for the asymptotic
expansions~\cite{Smirnov:2012gma} and {\tt MATAD}~\cite{Steinhauser:2000ry},
written in {\tt Form}~\cite{Form}, for evaluation of the three-loop diagrams.
\begin{figure}[t]
  \begin{center}
    \begin{tabular}{cc}
      \includegraphics[width=0.48\textwidth]{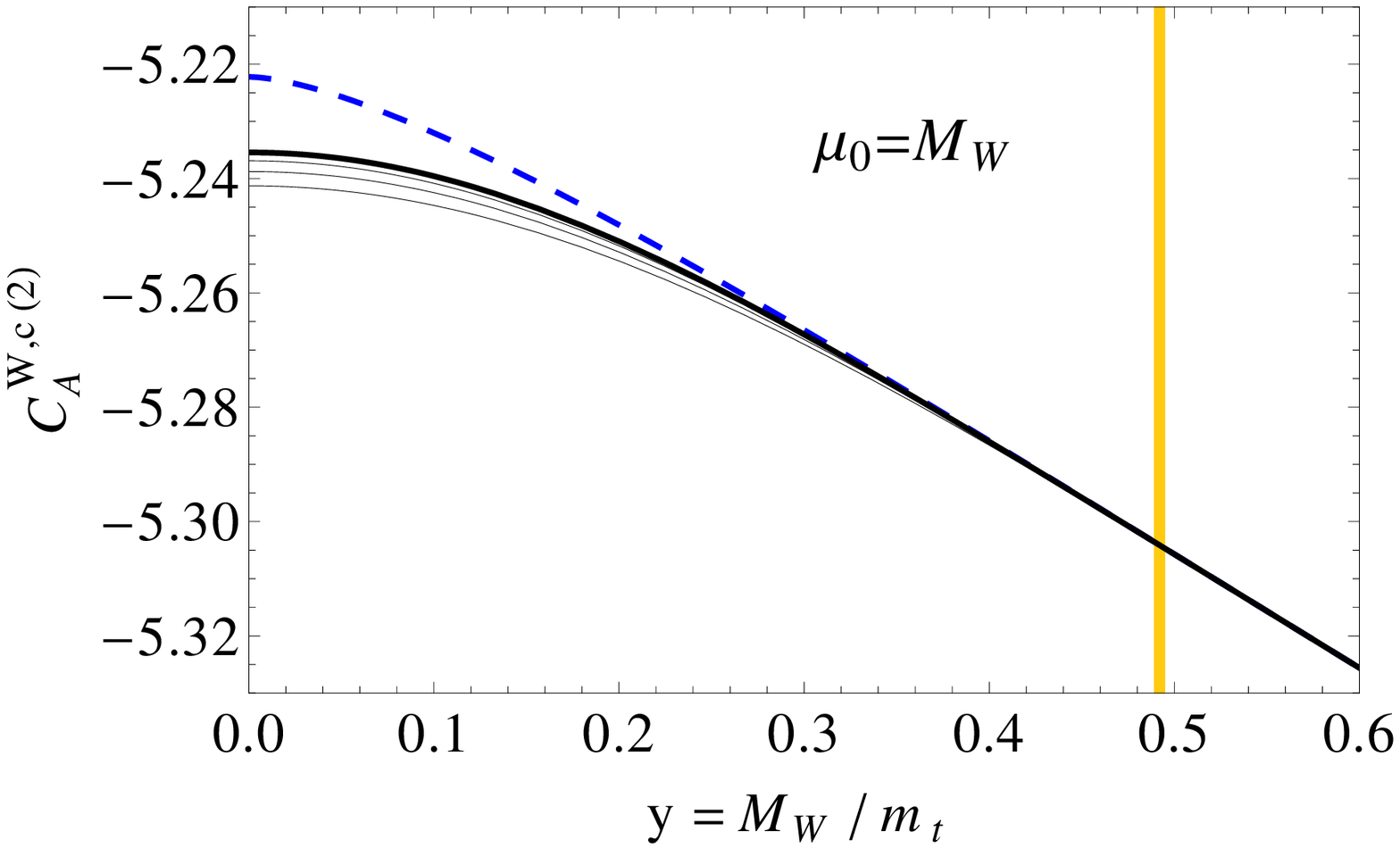} &
      \includegraphics[width=0.48\textwidth]{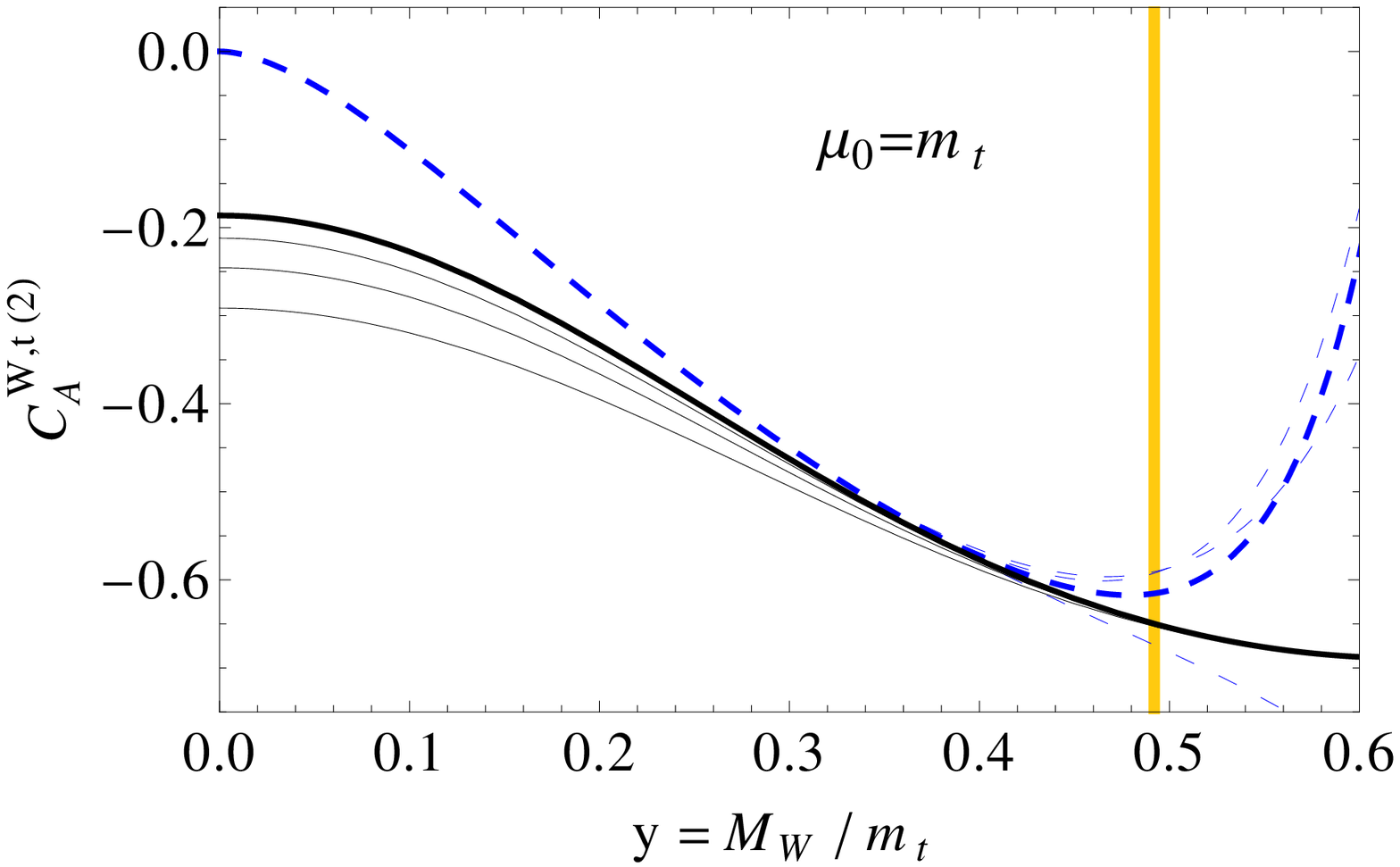}
    \end{tabular}
     \caption{
      \label{fig::result_W_box}
      $C_A^{W,(2)}$ as a function of $y=M_W/m_t$ for the charm (left) and
      top quark sector (right). The (blue) dashed lines are obtained in the 
      limit $y\ll 1$, and the (grey) solid line for $y \approx 1$.
      Thinner lines contain less terms in the expansions. The physical region
      for $y$ is indicated by the (yellow) vertical band.}
  \end{center}
\end{figure}

In Fig.~\ref{fig::result_W_box}, the results for the three-loop
corrections to $C_A^{W}$ are shown as
functions of $y=M_W/m_t$. The dashed and solid lines
correspond to the $y\to 0$ and $y\to 1$ expansions,
respectively. For the charm quark (left panel of Fig.~\ref{fig::result_W_box})
as well as for the top quark contribution (right panel of
Fig.~\ref{fig::result_W_box}), the two
different expansions show a nice overlap. Considering the thin lines in Fig.~\ref{fig::result_W_box},
which represent lower terms in the expansions, they indicate a good convergence for both expansions.
In the physical region of $y$ (yellow band), the expansion arround $m_t=M_W$
is sufficient.

The second type of diagrams contributing to $C_A$ are the Z-boson penguins, sample diagrams
are shown in Fig.~\ref{fig::z_penguin}.
For those contributions it is necessary to introduce
an electroweak counterterm.
In addition, at the three-loop level one encounters diagrams with triangle
quark loops that involve axial current couplings to the Z-boson.
In these cases, one needs to be careful about the treatment of $\gamma_5$.
In this work we follow the strategy of trace evanescent operators~\cite{Gorbahn:2004my}
and cross-checked it against Larin's method~\cite{Larin:1993tq}.
\begin{figure}[t]
  \begin{center}
    \begin{tabular}{cc}
      \includegraphics[width=0.48\textwidth]{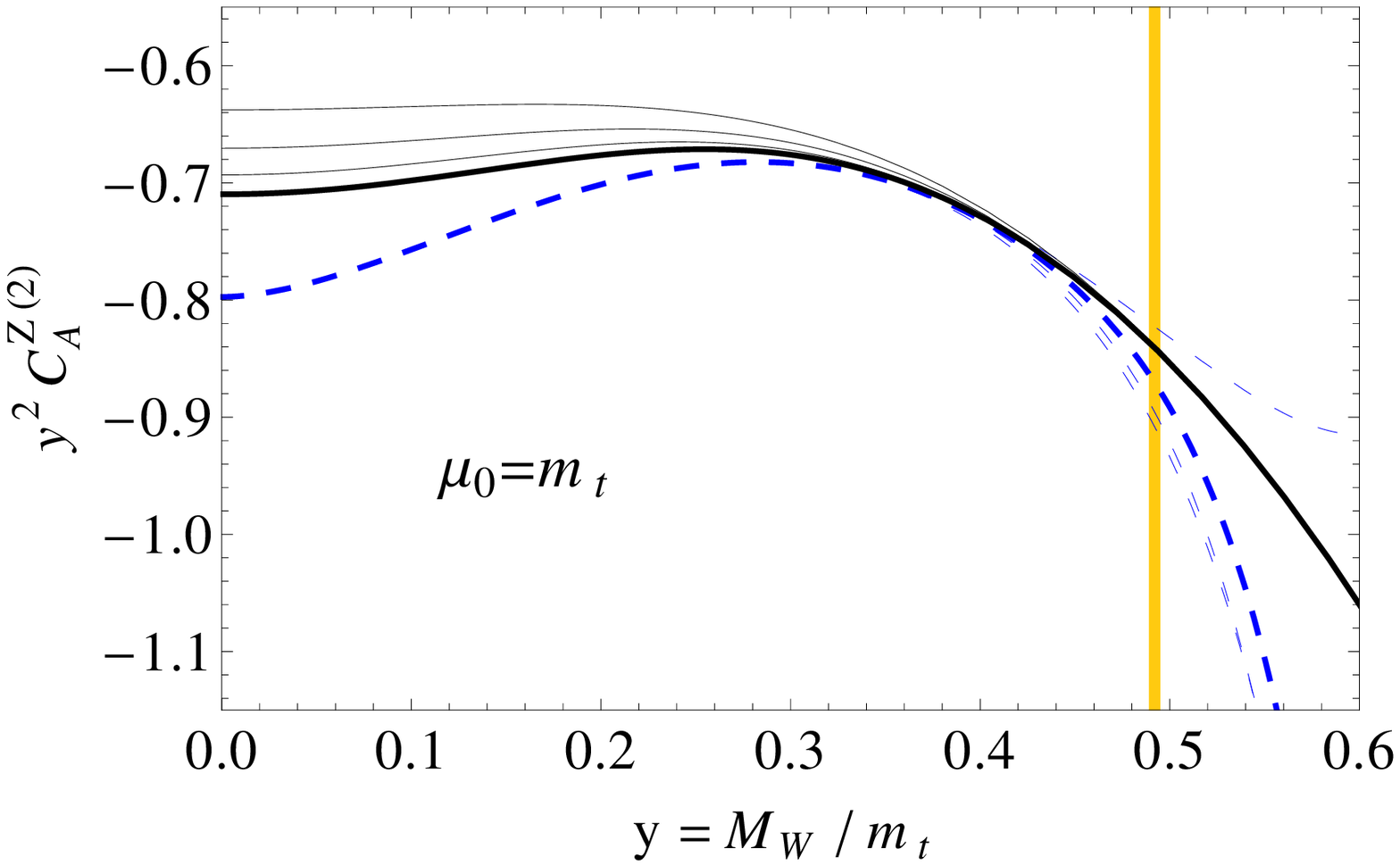} &
      \includegraphics[width=0.48\textwidth]{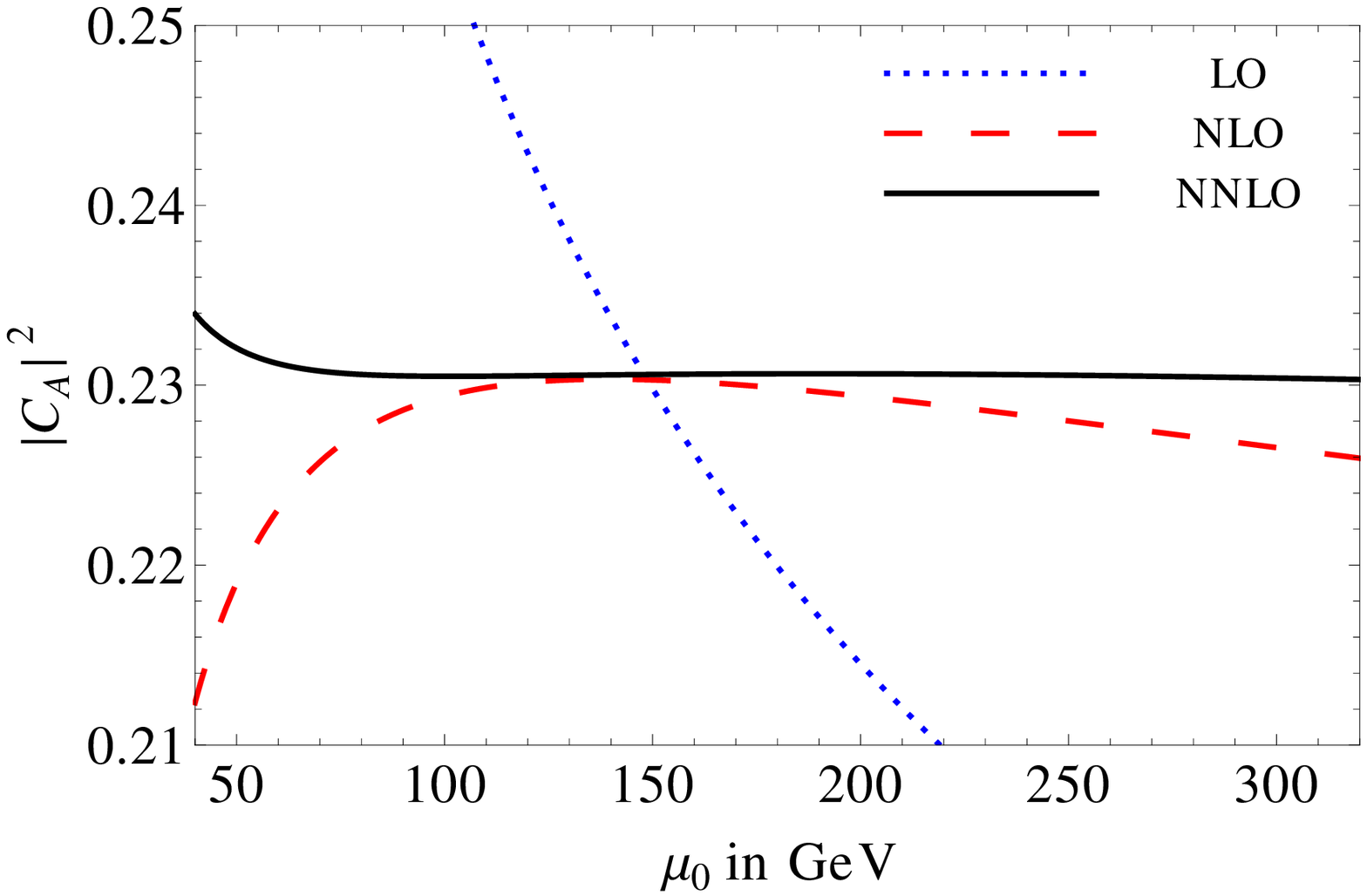}
    \end{tabular}    
    \caption{
      \label{fig::result_Z_penguin}
      Left: $y^2 \, C_A^{Z,(2)}$ as a function of $y=M_W/m_t$.
      Right: Matching scale dependence of $\left|C_A \right|^2$ at LO, NLO and NNLO in QCD
      without higher order corrections in EW interactions.}
  \end{center}
\end{figure}
In the left panel of Fig.~\ref{fig::result_Z_penguin} the sum of the three-loop contributions to $C_A^Z$ are shown.
The two expansions for $y\to 0$ (dashed line) and $y\to 1$ (solid line) again overlap.
In the physical region (yellow band), the expansion $y \to 1$ is again sufficient.

In the right panel of Fig.~\ref{fig::result_Z_penguin} the matching scale dependence of $|C_A|^2$
is shown. In the SM the branching ratio is proportional to $|C_A|^2$ (cf. Eq.~(\ref{eq::BR})).
The dotted, dashed and solid curves in the right panel of Fig.~\ref{fig::result_Z_penguin} show
the leading order (LO), next-to-leading-order (NLO) and the new next-to-next-to-leading order (NNLO)
results.
The variation of the matching scale for $\frac12m_t < \mu_0 < 2 m_t$ amounts to around $1.8\%$ at the NLO.
With the new three-loop QCD corrections the uncertainty gets reduced to less than $0.2\%$.

The results shown so far are at LO in EW interactions.
The combination with the recently calculated NLO EW corrections~\cite{Bobeth:2013tba} and
the RGE running from Refs.~\cite{Bobeth:2003at,Huber:2005ig} leads to the branching ratio
$\overline{\mathcal B}(B_s \to \mu^+ \mu^-) = (3.65 \pm 0.23) \times 10^{-9}$.
Details on the numerical analysis can be found in Ref.~\cite{Bobeth:2013uxa}.

\section{\label{sec::bsgamma}$\bar{B}\to X_s \gamma$ in the 2HDM at NNLO in QCD}

For ${\mathcal B}(\bar{B}\to X_s \gamma)$ in the 2HDM type-II we calculate
the three-loop QCD corrections to the Wilson coefficients of the operators
\begin{eqnarray}
Q_7 &=& \frac{e}{16\pi^2} m_b \left( \overline{s}_L \sigma^{\mu\nu} b_R\right) F_{\mu\nu}\,,\\
Q_8 &=& \frac{g}{16\pi^2} m_b \left( \overline{s}_L \sigma^{\mu\nu} T^{a} b_R\right) G^{a}_{\mu\nu} \nonumber \,.
\end{eqnarray}
The three-loop matching for $C_7$ and $C_8$ in 2HDMs works similar to the SM matching~\cite{Misiak:2004ew},
details of the calucaltion can be found in Ref.~\cite{Hermann:2012fc}.

The interaction Lagrangian for the charged Higgs boson with quarks reads
\begin{eqnarray}
  {\cal L}_{H^+} &=& (2\sqrt{2}G_F)^{1/2} \sum_{i,j=1}^3 \overline{u}_i \left(
    A_u m_{u_i} V_{ij} P_L 
    - A_d\, m_{d_j} V_{ij} P_R  \right) d_j H^{+} + h.c.
  \,.
\end{eqnarray}
In the type-II model the coefficients $A_d$ and $A_u$ are given by
\begin{eqnarray}
 A_u = -\frac{1}{A_d} = \frac{1}{\tan{\beta}} \,,
  \label{eq::typeII}
\end{eqnarray}
where $\tan{\beta}$ is the ratio of the vacuum expectation values of the two Higgs doublets.
\begin{figure}[t]
  \begin{center}
    \includegraphics[width=0.8\textwidth]{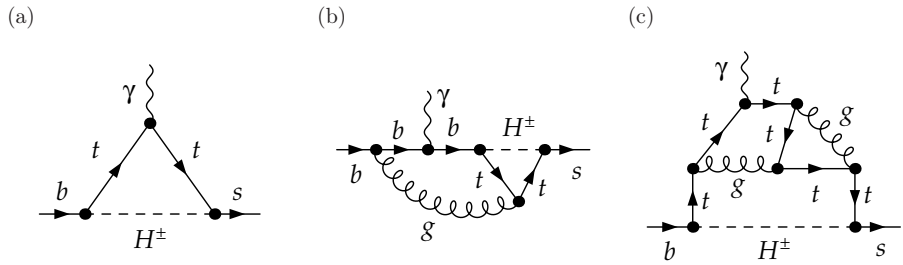}
    \caption{\label{fig::diags} Sample Feynman diagrams contributing to $C_7$
    at one-, two- and three-loop order.}
  \end{center}
\end{figure}
In analogy to the calculation of $C_A$, we have to consider vacuum integrals with two mass
scales ($\mhplus$ and $m_t$). Sample diagrams for $C_7$ up
to three-loops are shown in Fig.~\ref{fig::diags}.
At the one- and two-loop level, the calculation can be performed exactly, and one obtains 
$C_7$ as a function of 
$m_t/\mhplus$~\cite{Ciuchini:1997xe,Borzumati:1998tg,Ciafaloni:1997un,Bobeth:1999ww}.
At the three-loop level, we proceed as in Section~\ref{sec::Bsmumu} and
consider expansions around $m_t\approx \mhplus$, for $m_t\ll \mhplus$ and
for $m_t\gg \mhplus$.
\begin{figure}[t]
\begin{center}
  \begin{tabular}{cc}
    \includegraphics[width=0.48\textwidth]{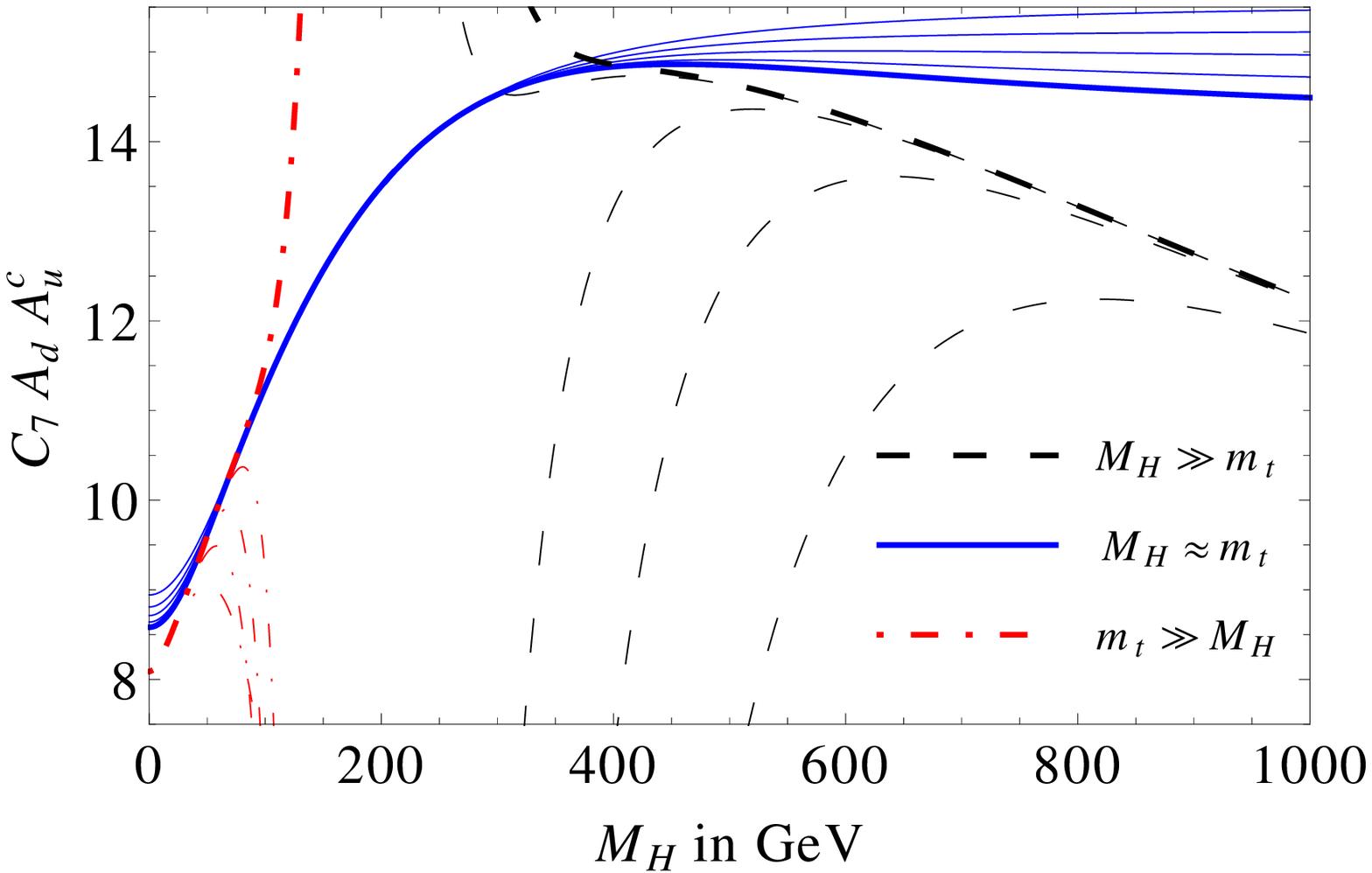} &
    \includegraphics[width=0.48\textwidth]{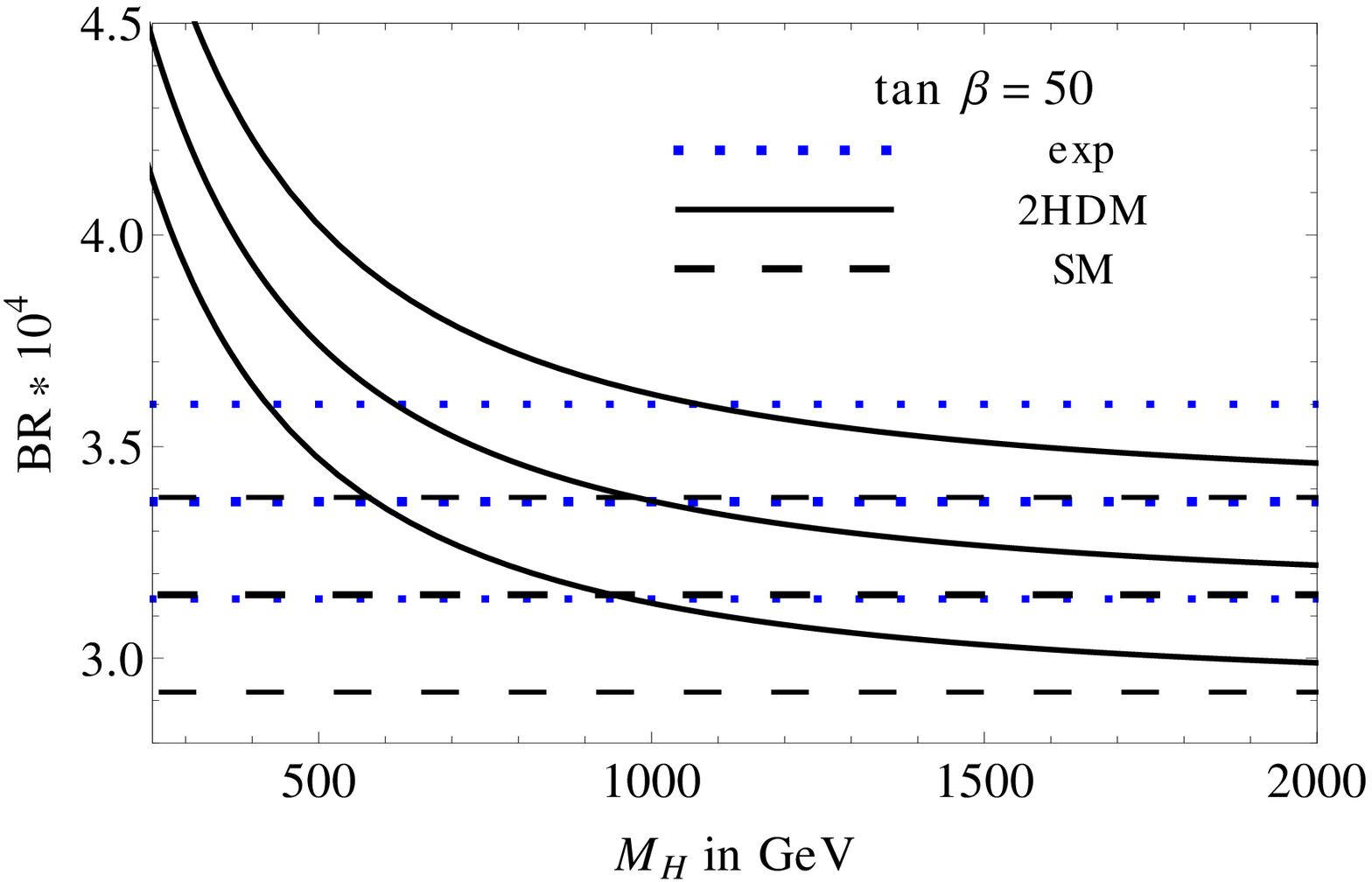}
    \\
    (a) & (b)
  \end{tabular}
    \caption{Left: Three-loop coefficient $C_{7,A_d A_u^{*}}$ as a function of $\mhplus$ for the three different expansions.
      Right: ${\mathcal B}(\bar{B}\to X_s \gamma)$ in dependence of $\mhplus$. Solid and dashed lines correspond to the NNLO 2HDM and SM predictions, 
      the dotted curves represent the experimental average, all with their respective $1\sigma$ uncertainty band.}
      \label{fig::2HDM}
  \end{center}
\end{figure}
In the left panel of Fig.~\ref{fig::2HDM} the part of the three-loop correction to $C_7$ proportional to $A_d A_u^{*}$
is shown in dependence of the charged Higgs boson mass. The thick-dashed, solid and dash-dotted lines
show the results for $\mhplus\to\infty$, $\mhplus\approx m_t$ and $\mhplus\to0$, respectively.
There is an overlap between the expansions for $\mhplus\to0$ and $\mhplus\approx m_t$ as well
as for the expansions $\mhplus\approx m_t$ and $\mhplus\to\infty$, which implies that the expansions
are sufficient to obtain $C_7$ for any value $\mhplus$. For the second part of $C_7$ (proportional to $A_u A_u^{*}$)
and $C_8$ the results look very similar.

For the calculation of the branching ratio we use all known NNLO QCD ingredients from Ref.~\cite{Misiak:2006ab,Misiak:2006zs}.
In the 2HDM the branching ratio depends on $\tan\beta$ and $\mhplus$. For the type-II model
the branching ratio is almost independent of $\tan\beta$, for $\tan\beta \gtrapprox 2$. For $\tan\beta \lessapprox 2$ the branching ratio
is strongly enhanced and much higher than the experimental results.
In the right panel of Fig.~\ref{fig::2HDM} the branching ratio is shown as a function of the charged Higgs mass for
$\tan\beta=50$ (solid lines). In addition, the SM prediction (dashed lines) and the experimental average (dotted lines)
are shown. The middle lines represent the central values, while the upper and lower ones are shifted by $\pm 1 \sigma$.
From Fig.~\ref{fig::2HDM} one can extract the following limit for $\mhplus$~\cite{Hermann:2012fc},
\begin{eqnarray}
\mhplus &\ge& 360~\mbox{GeV} \quad \mbox{at } 95\%~\mbox{C.L.} \, ,
\end{eqnarray}
where the experimental average ${\cal B}(\bar{B}\to X_s \gamma)|_{E_\gamma>1.6~\mbox{\tiny GeV}} =(3.43 \pm 0.22) \times 10^{-4}$
from the HFAG web page~\cite{hfag} has been used.

\section{Conclusion}
In this contribution the calculation of the Wilson coefficients $C_A$ in the SM and
$C_7$ in 2HDMs to NNLO in QCD were discussed. Inclusion of the three-loop corrections
leads in both cases to a significant reduction of the matching scale dependence.
Together with the NLO EW corrections~\cite{Bobeth:2013tba} for $C_A$, the SM prediction
is given by $\overline{\mathcal B}(B_s \to \mu^+ \mu^-) = (3.65 \pm 0.23) \times 10^{-9}$~\cite{Bobeth:2013uxa}.
In the 2HDM type-II a lower limit of $\mhplus \ge 360~\mbox{GeV}$ at $95\%~\mbox{C.L.}$~\cite{Hermann:2012fc}
for the charged Higgs boson mass has been obtained.

\section*{Acknowledgements}
I thank Miko{\l}aj Misiak and Matthias Steinhauser for the fruitful collaboration and
carefully reading this manuscript. I also thank Franziska Schissler for helpful comments.
This work was supported by the DFG through
the SFB/TR~9 ``Computational Particle Physics'' and the Graduiertenkolleg
``Elementarteilchenphysik bei h\"ochster Energie und h\"ochster
Pr\"azision''.

\end{document}